\documentclass[aoas,MSNbibl,nameyear,dvips]{arximspdf}
\usepackage{graphicx}
\doi{10.1214/13-AOAS677} 
\volume{7}
\issue{4}
\pubyear{2013}
\firstpage{2081}
\lastpage{2105}



\begin{document}
\begin{frontmatter}

\title{The tomb next door: An update to ``Statistical analysis of an
archeological find''}
\pdftitle{The tomb next door: An update to ``Statistical analysis of an
archeological find''}
\runtitle{The tomb next door}

\begin{aug}
\author[A]{\fnms{Andrey} \snm{Feuerverger}\corref{}\ead[label=e1]{andrey@utstat.toronto.edu}}
\runauthor{A. Feuerverger}
\affiliation{University of Toronto}
\address[A]{Department of Statistics\\
University of Toronto\\
100 St George Street\\
Toronto, Ontario\\
M5S 3G3, Canada\\
\printead{e1}} 
\end{aug}

\received{\smonth{3} \syear{2012}}
\revised{\smonth{8} \syear{2013}}

%
\begin{abstract}
In June of 2010
access via robotic means was obtained
to a tomb adjacent
to the one studied in Feuerverger [\textit{Ann. Appl. Stat.} \textbf{2} (2008) 3--54].
In this update, we lay out and attempt to interpret
the remarkable findings from this second tomb
and comment on the statistical and scientific significance
of these new data and of their possible inferential connections
to the data from the first tomb.
Readers are then invited to formulate their own conclusions.
\end{abstract}
%
\pdfkeywords{A priori hypotheses, Book of Jonah, coincidence, exploratory and confirmatory experiments,
ossuaries, resurrection, scientific inference, statistical inference}
\end{frontmatter}

\section{Introduction and summary} \label{Sect.Intro}

The purpose of this article is to update the discussion
in \citet{feuerverger} (hereafter AF08)
concerning a certain tomb (hereafter Tomb 1)
in the East Talpiot suburb of Jerusalem
in light of recent additional findings.
We refer the reader also to the Discussion,
as well as to the Rejoinder,
of the mentioned paper.

The tomb studied in AF08 contained ten ossuaries,
of which six bore inscriptions of names that,
while mostly common,
were reminiscent of the New Testament (NT) family.
In that paper, the archeological context,
background on the practice of ossuary interment,
the onomasticon of the era,
as well as some historical and genealogical information,
were laid out in some detail.
There, a ``historical'' approach was adopted
which, in particular, meant that the possible existence
of a NT tomb site in the vicinity of Jerusalem
was not viewed as being
implausible.\footnote{Although
there is a broadly-based social conditioning
that this is unlikely or impossible,
Jewish customs of the era are
consistent with the possible existence of such a tomb,
and it is known that Jesus died in Jerusalem and was,
on at least one occasion, buried in a tomb there.}
It was then computed---under various sets of assumptions
\textit{which are far from universally agreed upon}---that the probabilities
(under random assignment of names from the onomasticon)
of drawing a tomb site
as closely matching to the NT family
as the one at Talpiot
were typically less than one percent.
Such significantly small probabilities were driven heavily
by what was assumed,
from among the available names,
and on a presumed {a priori} basis,
to be the most fitting name for Mary Magdalene;
the results are not (statistically) significant without
that {a priori} assumption.
See, for example, the Rejoinder of AF08,
as well as the \hyperref[Sect.Appendix]{Appendix} to this paper.
The reader should, however, also be aware
that there are other controversies\footnote{In
assessing such controversies one does, however,
need to distinguish between arguments that have a rational basis
and arguments that do not.}
surrounding that analysis;
a representative selection of these
arise in the Discussion of AF08.
See also the Volume 69 issue of Near Eastern Archeology [Near Eastern Archaeology (\citeyear{Near})] and
Kloner and Gibson (\citeyear{klonerE}).
In the Rejoinder of AF08, the potential value of excavations
at an immediately adjacent tomb site was alluded to,
while noting also the strictness of Israeli laws
governing matters that pertain to disturbing burial sites.

At the end of June, 2010, however, R. Arav, S. Jacobovici,
and J. Tabor succeeded, after some efforts,
to obtain limited access to a tomb (hereafter Tomb 2)
adjacent to, and some sixty meters
to the north by northwest, of Tomb 1.
To obtain that access, it was first necessary to
(i) secure permission from the Research Department
of the Israel Antiquities Authority (IAA),
(ii) obtain an archeological excavation permit from the IAA,
(iii)~secure cooperation of local police
to deal with possible religious tensions,
(iv) obtain permission from the tenants association
of the condominium building situated above the tomb,
and (v) consult an engineer to assure
that the building would not suffer structural damage
when metal reinforcements in concrete
blocking access to the tomb were cut.
Obtaining consent from religious authorities
was an entirely separate matter involving
delicate negotiations with the spiritual leaders
of Bnei Brak-based ultra-Orthodox communities
adamantly opposed to any access or interference
in Jewish gravesites;\footnote{
The sensitivity of ultra-Orthodox Jews (Haredim)
to disturbance of burial sites,
and, in particular, to interference with the bones of the dead,
stems from a central belief of classical Judaism,
namely, that the dead would be raised
upon arrival of the Messiah.}
without that cooperation,
access to the tomb would not have been possible.

The agreements that could be reached
stipulated, in particular, that
the tomb could be explored,
but \textit{without any physical entry} into it,
and that no bones or ossuaries were to be touched.
Therefore, access was secured by means of a shaft,
bored through the floor of a corridor
of the apartment complex above the site.
For that purpose a 20 centimeters
diameter custom-made diamond tooth drill was used, together with
ground-penetrating radar to assist in determining where to drill.

Through this shaft, a reconfigurable, highly modular,
and pneumatically operated
multi-actuating extensible robotic arm---designed and custom built for this purpose\vadjust{\goodbreak}
by robotics engineer Walter Klassen---was inserted.
This arm was mounted with a small, waterproof
GE Inspection Technologies ``Pan till Zoom''
module with built-in halogen lighting
and high-definition camera.
Two smaller holes
were also drilled
to permit a secondary light source and a still smaller camera
to be inserted; their main purpose was to help guide
the manual remote control of the robotic arm.
By this means, an exploration
of the interior of the tomb,
and of the ossuaries located within it,
was carried out.\enlargethispage{3pt}

The archeological findings that were obtained in this way
are described in Section~\ref{Sect.Findings},
and some possible interpretations of those findings
are discussed in Section~\ref{Sect.Interpretations}.
Some background historical material
is postponed to Section~\ref{Sect.History}.
The nature of the new findings
raises nontrivial questions of statistical and of scientific inference;
our discussion of these is presented in
Sections~\ref{Sect.Data} and \ref{Sect.Inference},
the first of which deals with issues of data,
and the second with issues of inference,
after which readers are invited to form their own conclusions.
Some closing remarks are given in Section~\ref{Sect.Conclusion}.
Due to its inferential significance,
a further discussion concerning the interpretation
of the Greek ``Mariamne'' inscription of Tomb 1
and some relevant new data are provided in an
\hyperref[Sect.Appendix]{Appendix}.\vspace*{-1pt}

\section{The archeological findings} \label{Sect.Findings}

Whether laid out prior to contact with the data or
(on some ``best efforts'' basis) only afterward,
it is doubtful that anyone's set of {a priori} hypotheses
could have adequately encompassed the essence
of what was found in this adjoining tomb.

Its ceiling lay
below current ground level,
and a square \textit{golal} (sealing stone)
to its entrance was still in place.
Within it were scattered human skeletal remains
and three \textit{kokhim} (niches)
carved into each of three limestone walls---nine niches in all.
Within these niches,
a total of \textit{seven} ossuaries were found.
Six of the ossuaries were nicely decorated---typically with inscribed and colored
(yellow or red) circular rosettes;
the seventh could not be examined on all its sides
due to limitations of the physical setup,
but it appears to have been plain (undecorated).
Except for one unfinished rosette,
all of the mentioned decorative patterns
were rendered fully.
One of the ossuaries bore detailed images
and two of them bore inscriptions in Greek lettering.
These ossuaries, numbered arbitrarily here,
are described as follows:\looseness=-1

\begin{figure}

\includegraphics{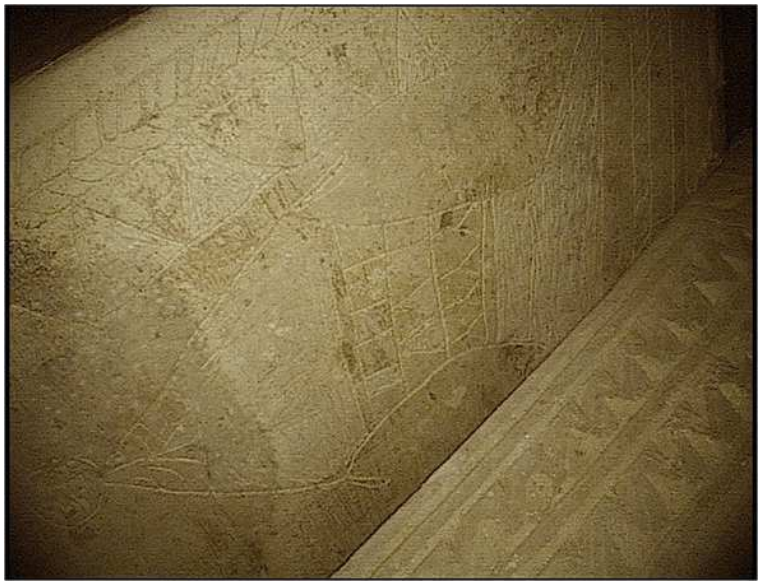}

  \caption{Front left half
of Ossuary \textup{\#1} (unenhanced photograph, rotated).}\label{fig1}
\end{figure}

\textit{Ossuary} \#1:  The front left half of this ossuary
is inscribed with an image that appears to be that of a fish pointing
downward.\footnote{The
front right half of this ossuary could be observed only partially;
it appears to consist of unobserved content, surrounded
by architectural bordering.}$^{,}$\footnote{The
``fish'' is scaled
and has proportionate, appropriately located fins
which narrow at the body.
It also appears to have both a head and an outwardly fanning tail.}$^{,}$\footnote{Note
to the reader:
That the image is that of a fish is \textit{far from uncontested},
and (\textit{qv}) neither is it inconsequential;
it has been suggested that it represents
either an amphora or unguentarium, or a stele
(although this ossuary does appear to contain other fish motifs on it).
\label{footnoteA}}
The mouth of the ``fish'' is closed,
with something round protruding from it.
This round object appears to be
marked up with interwoven lines.
Surrounding the front of this ossuary are smaller ``fishes''
and bordering.
The left side of this ossuary contains a
bordered cross-like image,
while the right side appears to show part of a ``fish,''
possibly diving.
This ossuary contains no rosettes or other decorative patterns,
and its back is plain.
An (unenhanced) photograph of the ``fish'' on the front of this ossuary
is shown in Figure~\ref{fig1};
in this photograph the ``fish'' has been rotated $90^\circ$ clockwise
and is shown facing leftward.

\begin{figure}

\includegraphics{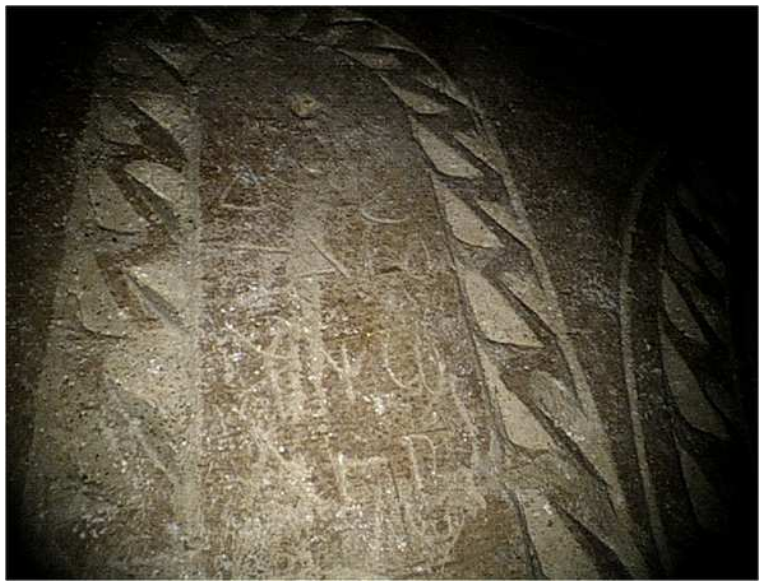}

  \caption{Front of Ossuary \textup{\#2} (unenhanced photograph)
showing four-line inscription between rosettes.
The fisheye distortion in this image
is due to narrowness of the passage
between that ossuary and the niche's wall.}\label{fig2}
\end{figure}

\textit{Ossuary} \#2:  The front of this ossuary is nicely ornamented
with two painted circular rosettes
between which appear four lines of text
informally inscribed in upper case Greek script.
The first three lines appear to read:\footnote{Note
to the reader: The readings indicated here are also
far from uncontested. See, for example, the detailed discussions
posted on the \citeauthor{ASOR} blog site. \label{footnoteB}}
\begin{eqnarray*}
&\Delta I  O  \Sigma,&\\
&I  A  I  O,&\\
&\Upsilon \Psi \Omega.&
\end{eqnarray*}
The fourth line is less clear and may be one of
\[
\mathrm{(a)}\quad  A  \Gamma B,\qquad    \mathrm{(b)}\quad  A  \Pi \Omega,\qquad   \mathrm{(c)}\quad  A  \Gamma I \Omega
\]
or some close variant.
The two sides and back of this ossuary are otherwise plain.
An (unenhanced) photograph of this inscription is shown in Figure~\ref{fig2}.

\begin{figure}

\includegraphics{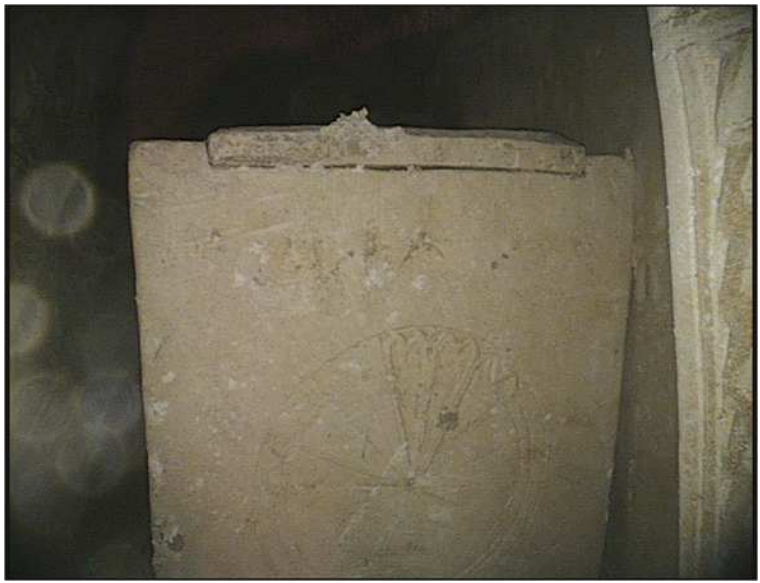}

  \caption{Ossuary from Tomb 1 with
the Greek inscription.}\label{fig3}
\end{figure}

\textit{Ossuary} \#3:  One side of this ossuary contains the inscription
\[
M A P A
\]
(i.e., Mara) in upper case Greek script
and an incomplete,
primitively rendered rosette.
The rest of this ossuary is plain.
See Figure~\ref{fig3}.

\textit{Ossuary} \#4:  This ossuary is nicely decorated
with a border design and two painted circular rosettes
carved on its front.
On its upper right corner is a stick-like image
in the shape of a short caterpillar
that could not be deciphered.
(Such images on ossuaries are rare.)
The remaining sides of this ossuary are plain.

\textit{Ossuary} \#5:  This ossuary is nicely decorated
with two well executed and painted circular rosettes
on its front surrounded by border design.
Between the rosettes is a symbolic pillar
known as a \textit{nefesh.}
The \textit{nefesh} symbol is discussed
in Section~\ref{Sect.Interpretations}.
The other sides of this ossuary are plain.

\textit{Ossuary} \#6:  Due to its positioning,
this ossuary could not be examined properly.
The ossuary was ornamented
and appears to have a Greek name inscribed on it
that could not be
read.

\textit{Ossuary} \#7:  This ossuary is completely plain.

\textit{Ossuary} \#8:  An eighth and child-sized ossuary that was not in the tomb,
but which is known to belong to it,
will be introduced in Section~\ref{Sect.History}
where its circumstances will be explained.
It is painted and nicely decorated,
with two symmetrically arranged circular rosettes
between which is carved a \textit{nefesh.}

For these and some further details about the finds,
see \citet{TaborPrelim} and \citet{TaborJacobovici}.

\section{Some interpretations} \label{Sect.Interpretations}

To understand the evidentiary value of these ossuaries
requires consideration and interpretation
of their inscriptions.
As a general remark,
it may be surmised that ossuaries decorated as nicely
as some of these are
would typically be thought of
as having belonged to more well-to-do families.

We begin with the remarkable Ossuary \#1,
the one with the image of what may be a fish.\footnote{See,
however, footnotes \ref{footnoteA} and \ref{footnoteB}.}
If it is, then it appears to be that
of an eastern\footnote{Israeli
specialist of ancient art history, Shua Amorai-Stark,
points out that ancient images of western fish
are usually easily identified as to species,
while eastern fish, like this one,
tend to be more abstract [private communication].}
fish, reminiscent of the ``big fish''
[\textit{dag gadol}---sometimes translated as ``great fish'' or ``whale'']
in the Book of Jonah.
This ``fish'' has scales and fins,
consistent with the requirements of kashruth.
As its mouth is closed,
it is not evidently in the process of swallowing anything,
but might instead be spewing something out.
The round object outside its mouth
is proportionate to a human head,
and the carvings---which appear purposeful---that mark up this ``head''
are consistent with seaweed-like material.
In the book of Jonah, Chapter~2:6, there appears the line:
``The engulfing waters choked me, the deep surrounded me,
seaweed was wrapped about my head.''\footnote{This
theme is echoed in Jonah, Chapter~4:6,
``And God created a \textit{kikayon},
and made it to come up over Jonah,
that it might be a shadow over his head,
and save him from evil.''
The exact modern translation for the fast-growing
\textit{kikayon} plant is uncertain,
and it is commonly mistranslated as ``gourd;''
see, for example, \citet{janick}.}
Attached to this ``head''
and inside the mouth of the fish
appears to be the (stick-like) remainder of a human body.
The body is rendered primitively,
consistent with the stone medium and with
an observation of Jensen [(\citeyear{jensen}), page 12] that
``the earliest examples of Christian art
are simple, almost humble, in their manner of presentation.''

Within Christianity, the story of Jonah is commonly
interpreted as a story about death and resurrection---a quintessentially Christian theme.\footnote{For
Jews, the story of Jonah represents the notion
that no one is beyond divine forgiveness;
it has a much lesser messianic significance.}
As Jensen [(\citeyear{jensen}), page 51] puts it:

\begin{quote}
``Jonah, especially, serves the double function
of symbolizing both Christ's death and his resurrection---the ``sign'' of Jonah (Matthew 12:39 and parallels),
and the baptism of each believer.''
\end{quote}

\noindent However, as a symbol of Christianity the fish is not known
to have appeared until a significantly later time.
Jensen [(\citeyear{jensen}), page 9] states:

\begin{quote}
``Christian art as such cannot be dated any earlier
than the end of the second or beginning of the third century.
Before that date, material evidence of Christianity is scarce
and, although not entirely nonexistent, often hard to distinguish
from objects that belonged to the wider cultural context.''
\end{quote}

\noindent
Quoting further from \citet{jensen}:

\begin{quote}
(page 21): ``Almost all existing pre-mid-fourth-century
art work was specifically created to decorate tombs or coffins.''\\
(page 172): ``The figure of Jonah was by far the most reproduced
in early Christian art. \ldots
In the pre-Constantinian era \ldots Jonah occurs more than seventy
times\ldots''\\
(pages 68--69): ``\ldots the story of Jonah is an overwhelmingly favorite subject\ldots
Slightly under one hundred Jonah figures are found in the catacombs
or carved in sarcophagi dated to the pre-Constantinian era alone.''
\end{quote}

\noindent
Snyder [(\citeyear{snyder}), page 54] too writes:

\begin{quote}
``The scene most used by early Christians was the Jonah narrative.''
\end{quote}

\noindent
And, finally, liturgical scholar Seaslotz [(\citeyear{seasoltz}), pages 115--116] writes:

\begin{quote}
``The figure of Jonah was one of the most frequently reproduced images
in early Christian art.
He is frequently shown being tossed into the sea,
being swallowed up by the fish,
emerging on dry land,
sitting under a gourd vine,
and as one who has come to life\ldots
Underlying the images is surely the theme of resurrection,
and it is directly linked with the text in Matthew 12:39--40
in which Jesus says that just as Jonah was three days
and nights in the belly of the fish,
so will the Son of Man be three days and nights in the earth.
\ldots the baptismal connection between Jonah and Jesus would be logical.
Jonah and the initiate are both immersed in water.''
\end{quote}

\noindent
Ancient symbols of Christianity may be found, for example,
in the catacombs under the streets of Rome.
There, the three most common among the earliest
Christian symbols include:
the fish (with scales and gills) as a symbol of Christ;
the anchor as a symbol of faith;
and the overlaid $\chi$-and-$\rho$
formed from the first two letters
of \textit{Christos} in Greek.
Of these symbols,
the fish is the one most commonly found.
However, those symbols in the catacombs date to the fourth century
or to the third century at the earliest.
No uncontestably Jesus-related artifacts
from the first century have ever been found,
and it is not known what symbols may have been used
by the very first followers of Jesus.

Noteworthy about Ossuary \#1 is that symbols of fish on Jewish
funerary objects are virtually nonexistent.\footnote{A
crude image of what may be a fish---without anything protruding from an open and upward-pointing mouth---does, however, appear on the ossuary of ``Claudius;''
see Rahmani [(\citeyear{Rahmani}), Item~348].
As for nonfunerary Jewish art,
graven images are extremely rare.
One image of a fish did occur on a table;
see Avigad [(\citeyear{Avigad}), illustration 185:4].}
Indeed, the use of ``graven images''
is forbidden in Jewish tradition,
so to find such an image in a Jewish tomb
of the first century is
quite unexpected.
This (together with the implication
of the scales and fins)
suggests that this ossuary
is associated with a Jewish person
who had transitioned away
from prevailing Jewish traditions.
Could then the symbol of a fish have been used
by the very earliest followers of Jesus,
and have \textit{subsequently}
found its way into wider use?
\textit{If so}, this would constitute the earliest
\textit{iconological} evidence
for the belief in resurrection ever found---and found on an ossuary box known to date
to within decades of Jesus' death.

The bordered cross carving on one side
of this ossuary may or may not be significant;
we attach no evidentiary value to it.
Cross-marks have appeared on other ossuaries,
either as mason's marks or as larger decorations.\footnote{The
``cross'' is generally not believed
to have become a symbol of Christianity
in the first century, but only at a later time
(although this point is not without some recent controversy).}

We next consider Ossuary \#2 and the four lines of text
that appear on it.
Although these four lines are executed in Greek script,
they appear to involve two languages---Greek and Hebrew.\footnote{The
reader is again referred here to footnotes
\ref{footnoteA} and \ref{footnoteB}.}
The first line, $\Delta I O \Sigma$, is Greek
and essentially refers to ``The Divine One'' or to ``God.''
The second line, $I A I O$,
is a transliteration into Greek script
of the Hebrew word for God as it appears
in the Old Testament: ``YHWH'' or ``Yehovah.''
Taken together, the first two lines
may constitute an address to Yehovah,
using the Greek and Hebrew languages alternatingly.
It is important to appreciate that the name Jehovah (YHWH) of God
\textit{never} appears on Jewish funerary objects;\footnote{This
is because death is associated with \textit{tuma}
(i.e.,  ritual impurity) and
God's name is \textit{never} placed onto anything impure.}
such an inscription constitutes
a \textit{very} significant violation of Jewish traditions,
again evidencing a departure from norms of the era.
Furthermore, citing the name of God \textit{twice}
in succession in this way symbolically violates
the worshipping of only \textit{one} god.

The third line, $\Upsilon \Psi \Omega$ ( ``UPSO''),
is in the Greek language
and unambiguously refers to the act
of ``raising'' or of ``lifting up.''
It could mean ``has raised,'' ``will raise,'' or ``is raising.''
The Greek verb, $\Upsilon \Psi \Omega$,
is used some 20 times in the New Testament,
including in the gospels of Matthew, Luke, and John,
as well as elsewhere.
For example, John 3:14 reads:
``And as Moses \textit{lifted up} the serpent in the wilderness,
so must the Son of Man be \textit{lifted up;''}
John 8:28 reads:
``When you have \textit{lifted up} the Son of Man,
then you will know that I am he;''
and John 12:32 reads:
``And I, when I am \textit{lifted up} from the earth,
will draw all people to myself.''
These uses of $\Upsilon \Psi \Omega$
pertain to the resurrection of Jesus.

The fourth line of the inscription
appears to consist of three characters of which the last two
are difficult to read.
If in Greek, it could mean ``the holy one'' ($agios$)
or ``the holy place'' or ``from death's realm.''
If in Hebrew (though in Greek script),
it could have the same meaning
as the third line, since \textit{hagbah} is the Hebrew imperative for
``lifting.''
In the latter case, the inscription
involves both Greek and Hebrew and reads:
``God, Yehovah, Raise up, Raise up.''

Among other possibilities
is that the fourth line was meant to be
the name of a person\footnote{As
a name it is very rare;
however, the name Agabus occurs in the Book of Acts (11:28 and 21:10).}
(Agba or Agaba).
A further possibility is that the second line
of the inscription was meant to refer to Jesus.\footnote{For instance,
John 10:33 states:
``\ldots because that thou, being a man, makest thyself God.''}
Professor James Charlesworth,
a specialist in New Testament languages and literature
at the Princeton Theological Seminary,
reads the four-line inscription as a plea to Jehovah
to lift or to raise someone up from the dead.\footnote{A
still further possible reading is:
``I wonderous YHVH, raise up, raise up.''}
Either way, it appears, in sum, to be a plea for resurrection.
If these readings are correct, then---because ossuary burials ceased in 70
CE---this would represent the earliest \textit{statement}
referring to resurrection ever found.
Furthermore, if one accepts the interpretations outlined here,
the implied meanings of Ossuaries 1 and 2
(found in the same tomb) are mutually reinforcing.

Ossuary \#3 bears the inscription MAPA (MARA),
which may be a shortened from of Martha,
although other possible interpretations
for this name-form were detailed in AF08.\footnote{It
may, for instance, have been a title.}
Because this name on ossuaries is rare,
its appearance in both of the adjacent Tombs 1 and 2
evidences a possible link between the families involved.

Ossuary \#8 was likely that of a child.
Inscribed between its circular rosettes
is a type of pillar commonly known as a \textit{nefesh};
the word \textit{nefesh} means ``soul,''
however, the image is symbolic of a monument or a stele.
Its occurrence on ossuaries is not rare.
This symbol is thought to have been adopted
from the Syrians and/or the Nabataeans
who viewed it as a dwelling-place
for the spirit after death
(in lieu of an actual monument),
and it was adopted by Jews who may have given it
a new meaning.\footnote{S. Jacobovici [private communication]
posits that the \textit{nefesh} symbol
may be a coded reference to the afterlife
symbolizing the ultimate resurrection of the dead.}
The \textit{nefesh} symbol is discussed in detail
in Hachlili [(\citeyear{Hachlili.three}), Chapter~8].

In sum, the ossuaries in Tomb 2, when taken together,
and when viewed in the context of the adjoining Tomb 1,
constitute an archeological find of considerable importance.
It is a find that has the potential to challenge the interpretations
previously assigned to numerous other archeological artifacts
unearthed over the years from the ancient city of Jerusalem.

\section{Some historical matters} \label{Sect.History}

The matters we deal with here refer to recent,
not to ancient, history.
Tomb 1, discussed in AF08,
was discovered on March 28, 1980
as a result of construction activity.
The site was visited by district archeologist
Amos Kloner on March 29th,
and salvage excavations (lasting a few days)
were begun on March 30th (under IAA permit 938) by Yosef Gath
of the Department of Antiquities and Museums,
assisted by Elliot Brown.
Shimon Gibson surveyed the site and drew up its plan.
See \citet{gath} and \citet{Kloner.three}.

The adjoining Tomb 2, described in this paper,
was actually first discovered in April, 1981,
also as a result of construction activity.
The Israel Antiquities Authority was notified,
and Kloner and an assistant were dispatched to investigate.
Through a hole inadvertently blasted in its ceiling,
the archeologists were able to descend
into this $3.5 \times 3.5$ meters tomb.
The tomb's entrance (nowadays some four meters below ground level)
was blocked by a square \textit{golal} (a large heavy sealing stone);
the other three walls each had three gabled niches carved into them,
which also were all blocked with stones.
Inside four of these niches,
a total of \textit{eight} ossuaries were found,
as well as some skeletal remains.
However, after only a short time in the tomb,\footnote{It
is uncertain now whether this was only for a few minutes,
a few hours, or a few days.}
the archeologists were set upon by an ultra-Orthodox group\vadjust{\goodbreak}
intent on preserving the sanctity of the site
and were forced to leave,
having had just enough time to
draw a rough map of the tomb's interior
and to take a few brief notes and a few
wide-angle black and white photographs
before the tomb was resealed.
Nevertheless, in the melee,
Kloner managed somehow to carry away
a small, uninscribed, but nicely decorated ossuary.
That ossuary is now housed at an IAA warehouse
in Bet Shemesh under catalogue number 81--505.
It is the one referred to
in Section~\ref{Sect.Findings} as Ossuary \#8.

Soon afterward,
IAA archeologists Yosef Gath\footnote{We
note that Yosef Gath was the
lead archeologist at \textit{both} of the tombs.}
and Shlomo Gudovitch,\footnote{When
recently interviewed, Gudovitch did not recall
any specifics concerning this excavation
beyond what is described here.}
with
permits secured, visited the tomb
for a period of several days.
They removed the (remaining seven) ossuaries from their niches
and recorded that all of them were decorated
and that two had Greek names inscribed on them.
However, \textit{no details of these inscriptions
\textup{(}and no mention of any images\textup{)}
were noted in their report,}
possibly on account of how little time
was available to them in the tomb.
In fact, just before the remaining seven
ossuaries could be hoisted away,
the archeologists were set upon by ultra-Orthodox activists
who insisted that the bone boxes
be put back into the niches.
In the circumstances,
these replacements were done haphazardly,
and the tomb was sealed on April 16, 1981.
Fortunately, however, the original placement
of the bone boxes had already been recorded
and is known;
see \citet{klonerB}. 
In particular, the ``big fish'' ossuary
is known to have come from the first niche
at the right of the tomb's entranceway,
the niche typically reserved for the patriarch of the family,
while Ossuary \#2 had been located in a different niche.

Yosef Gath died
in 1993
before having published his findings\footnote{Rahmani's
(\citeyear{Rahmani}) catalogue includes the ossuaries of Tomb 1
(acknowledging permission from Gath to publish them),
but not of Tomb 2, likely because its ossuaries did not find
their way into the State collections.}
on Tomb~2.
Partial reports
exist in unpublished archives
of the Israel Antiquities Authority
and, in particular, in
internal IAA memos dated April 17, and August~2, 1981 [Israel Antiquities Authority
(\citeyear{IAA})].
These reports are \textit{very} brief
and contain \textit{very} limited information.\footnote{\textit{Subsequent}
to the robotic exploration,
an obscure, never before cited clipping
dated 22 May 1981 (in Hebrew) from Davar---an Israeli newspaper that ceased publishing
well before such materials were put online---was unearthed.
The article's focus was on the Haredim interfering with the dig;
it only briefly mentions that there were
architectural images and a vase.}
Two reports on this tomb find were subsequently published:
\citet{klonerA} and \citet{klonerB}.
Concerning the ossuaries that were found in the tomb,
\citet{klonerA} states \textit{only} that:

\begin{quote}
``With the exception of one,
all of the ossuaries in the cave were decorated
with red or yellow paint or with incised designs,
including architectural facades.
Two of the ossuaries bore names incised in Greek.''
\end{quote}

\noindent
Except for the robotic entry of June, 2010,
the interior of this tomb has not been seen since.

It should be mentioned that there was also a third tomb
in the immediate vicinity of Tombs 1 and 2,
located approximately 20 meters north of Tomb 1.
Unfortunately, however,
that tomb was inadvertently completely destroyed during
construction activity in the area
and no record of its contents is available.
It is, of course, also possible that there may be other
undiscovered tombs in the region.

Finally, we include here a brief update
on the Aramaic-inscribed
``James son of Joseph brother of Jesus'' ossuary.
The trial of its owner, Oded Golan, accused of forgery,
which began at the end of 2004, ended in October, 2010,
and even though it is a norm
to be rendered within 30 days after a trial ends,
it was only in March 2012 that a verdict was handed down---a delay of unprecedented duration.
While this case was complicated by extraneous considerations,
the Court's decision was that it could find
no evidence that the ``James ossuary'' and its inscription were fake,
despite very intensive scientific examination
of that ossuary by experts.\footnote{Also
at issue was whether or not the ossuary
was acquired prior to 1978;
after that date its purchase would have been deemed illegal
so that ownership of the ossuary would transfer to the IAA.
Needless to say, if authentic, this antiquity would be priceless.}
The Court, however, also made it a point to say
that this did not mean that the inscription
has been proven authentic.
See, for example, \citet{golan}.
Cotton et al. [(\citeyear{cotton}) item 531, pages 547--548,
by J.~Price and A. Yardeni] comment as follows:
``Ossuary of Ya'akov son of Yosef brother of Yeshua
with Aramaic inscription, 1~c. BCE - 1~c. CE:''

\begin{quote}
``The origins of this ossuary can be traced no further
than its possession by a Tel Aviv antiquities collector,
who claims he purchased it in Jerusalem in the 1970s.
\ldots
The letter-forms seem appropriate for the first century CE
and cannot be decisively impugned on palaeographical grounds,
although all or parts of the inscription
(particularly the last two words) have been challenged\ldots
Authenticity is also disputed on the basis of petrographic analysis
of the patinas on the surface of the box
and within the grooves of the inscription (Ayalon et al.).
Yardeni and Lemaire have argued for its authenticity.
Yet even assuming it is entirely genuine,
the last two words, ``brother of Yeshua,'' would\footnote{The
use of the word ``would'' here, instead of ``might'' or ``could,''
evidences some degree of conviction
on the part of the authors.}
have been added to the normal name${} + {}$patronym
not because ``the brother had a particular role\ldots''
but\ldots to distinguish this Ya'akov
from a relative with a similar or identical name\ldots
Moreover, the grammar of the inscription allows Yeshu'a
to be the brother of either Ya'akov or Yosef:
there is no way of knowing.
\ldots
Context was completely lost when the object was looted from its cave.''
\end{quote}\vfill\eject

\section{Statistical issues: Data} \label{Sect.Data}

The questions that concern us next relate to the role that statistics,
as a discipline, might or might not be able to play
in analyzing and interpreting the findings
unearthed in Tomb 2,
when taken in conjunction with the data obtained from Tomb 1.
We begin with a discussion on available sources of data.

The first items of data pertain to the onomasticon---the names of the men and women
who lived during the era in question.
The study in AF08 relied on three sources:
(i) \citet{Rahmani}, who catalogued the ossuaries
in the collections of the State of Israel as of 1989,
of which some 233 bore inscriptions,
(ii) Tal Ilan's (\citeyear{Ilan.two}) lexicon of Jewish names
in late antiquity
that contains some 2826 names
when fictitious ones are excluded, and
(iii) \citet{Hachlili.three} that contains a subset
of Tal \citet{Ilan.two} dating to the late Second Temple period.
We point out that there is now
a recent, comprehensive
fourth source of such data,
namely, the nine-author edited volume
of \citet{cotton}.

The data in Cotton et al. are important for a number of reasons.
First, this source provides
a much more comprehensive collection of names
found on ossuaries than does any other.
Specifically, 591 funerary inscriptions are provided
[Cotton et al. (\citeyear{cotton}), entries 18 to 608],
virtually all of which are inscriptions taken from ossuaries.
This provides a much larger sample of such names than
hitherto available---almost three times as many as in \citet{Rahmani}.
No statistical summary of the names is provided,
but such a summary could be prepared by one so inclined.
Broadly put, there does not appear to be evidence here
that would substantively invalidate the frequencies of the names
occurring in the three mentioned earlier sources.

Second, Cotton et al. provide pictures
for a substantial proportion
of the inscriptions, and this is important for two reasons.
First, such pictures allow us to gauge the ``quality''
of ornamentation on typical ossuaries,
and hence to assess the ornamentation found in Tomb 2.
Loosely put, there is some small (although not unduly small)
proportion of ossuaries having ornamentation
as ``nice'' or ``nicer'' than was found in Tomb 2, that is, most of the
ossuaries in Tomb 2
belong to the category of ``nicer'' ossuaries.
Second---and this is particularly important---although the pictures in Cotton et al. were not intended
to focus on images (but rather on written inscriptions),
it is nevertheless clear from those many pictures
that Jewish funerary art
of the era explicitly excluded images of animate objects
and, particularly, any references to Yehovah (YHWH).
There is, in Cotton et al., not a single image
of any animal or of any person evident
on any of the many ossuaries illustrated there,
nor is there any reference to any Hebrew word for God.\footnote{The
publication of Cotton et al. preceded the robotic
exploration of the second cave.}$^{,}$\footnote{Note
that images of ossuaries may also be found
in \citet{Rahmani} and in \citet{Hachlili.three}.}
This corroborates the fact that graven images were forbidden,
in accordance with the Second Commandment.
It also lends some credence to the argument
that Ossuary \#1 (with the ``fish'')
and Ossuary \#2 (with the four-line inscription)
were associated with persons
having a decidedly different attitude
toward these prohibitions.

The following quotes from
Cotton et al. (pages 8--10, by B. Isaac)
are also relevant here:
\begin{quote}
\begin{longlist}[(a)]
\item[(a)] ``\ldots the overwhelming majority of known ossuaries
come from Jerusalem and its environs.''

\item[(b)] ``The expense involved in the excavation of the cave
and manufacture of the ossuary
would have favored people with more substantial means\ldots'' and:

\item[(c)] ``Of the ossuaries recorded to date,
only about 600--650 are inscribed,
and most of these inscriptions
only identify the name(s) of the deceased.''
\end{longlist}
\end{quote}

\noindent
It is worth mentioning that Cotton et al.
(entries 473--478, pages 495--501, by J.~Price and H. Misgav)
include in their volume an analysis of the burial cave
with the six inscribed ossuaries of Tomb 1. They state:

\begin{quote}
``Neither the entrance to the cave nor any of the loculi was found
sealed, and the
excavators noted signs of disturbance and looting before their arrival,
perhaps in
antiquity. Moreover, the cave was first inspected thoroughly by
excavators, and the
ossuaries removed, on a Friday; when they returned the next Sunday they
discovered
that the local residents had entered the cave and removed some of its contents,
including bones (Gibson). Much original data, including the original
placement of
the ossuaries in the tomb and their contents, were lost as a result of
the hurried pace
of excavations and disturbance by local residents, the untimely death
of the original
excavator (Gath) and his failure to keep detailed notes, and the
disturbance by
looters before the modern excavations. Sixteen years passed between the original
excavation by Gath and Kloner's final report; by that time the bones recovered
from the cave had been reburied without proper analysis.''
\end{quote}

\noindent
Cotton et al. were aware of media activity generated
by that first tomb and further state:

\begin{quote}
``If not for the coincidence of some of the
inscribed names with the central family of
the New Testament, this ordinary cave
and its unexceptional ossuaries should have
attracted little popular attention.
\ldots
There is no sound reason to connect any ossuary in this tomb
to any known historical figure.''
\end{quote}

\noindent
Cotton et al. also provide detailed analyses
for each of the six Tomb 1 ossuary inscriptions,
citing studies in peer-reviewed journals.
We shall return to their analysis of the
``Ossuary of Mariam(e) with Greek inscription''
in the \hyperref[Sect.Appendix]{Appendix} below.
In keeping with the scholarly objectives of their volume,
the analyses provided by Cotton et al.
maintain exemplary reserve.
We add here only that their work
predates the findings from the second tomb.

In our discussion of the aforementioned sources of data,
the implicit sampling unit, so far,
has been the individual ossuary.
There is, however, another relevant sampling unit,\vadjust{\goodbreak}
namely, the individual tomb.
Cotton et al. (pages 8--9) indicate that
to date some 900 tombs have been explored.
The data in the four already mentioned sources
are not summarized by tomb, but, here again,
such summaries could be prepared
from these sources by one so inclined.
A tomb-by-tomb itemization is, however, now available
in \citet{klonerC}.
In this reference, all tomb sites known as of 2002---a total of some 927 tomb
sites---are organized by zones (approximately 30 regions).
The tomb sites are described in varying levels of detail,
and references to published sources are given for each.
What is amply evident from all of these references,
however, is that tomb sites that provide so strongly
Judeo-Christian a message as Tomb 2 appears to do
are considerably more rare than 1 in 100---a fact that plays some role in the section
on inference below.

We mention that the compendium
of \citet{klonerC} includes
the two tombs that concern us here.
Tomb 1 appears in that reference
as item 12--46 [Kloner and Zissu (\citeyear{klonerC}), pages 342--343]
with the following remarks:

\begin{quote}
``Ten ossuaries, some decorated, were found in the cave
and its \textit{kokhim.}
Names such as ``Yehuda son of Yeshua''
``Matya,'' ``Yose,'' ``Marya,''
and ``Yeshua son of Yehosef'' were inscribed
in Hebrew on some of them.
Another ossuary belonged to
``Mariamene, (who is also called) Mara'',
inscribed in Greek on its long side.''\footnote{We
have quoted \citet{klonerC} here verbatim,
including not only their exact punctuation, but, in particular,
their interpretation of the Greek inscription as
``Mariamene (who is also called), Mara.''}
\end{quote}

\noindent
Tomb 2 appears in that reference
as item 12--45 [Kloner and Zissu (\citeyear{klonerC}), page~342] with the following remarks:

\begin{quote}
``A burial cave was discovered in the course
of development work,
and briefly examined by Kloner on behalf of the IDAM.

\ldots eight decorated and painted ossuaries
were found in the \textit{kokhim.}
Greek names were inscribed on two of the ossuaries.
Only one ossuary was removed from the cave.''
\end{quote}

\noindent
Gath's investigation of this tomb in not mentioned,
and Kloner and Zissu provide no other pertinent details
regarding the tomb.

We turn next to elements of data that pertain to images of fish.
Concerning such images,
the reign of Constantine\footnote{Constantine the Great,
commonly estimated to have been born in 272,
was Roman Emperor from 306 until his death in 337,
and the first Roman emperor to convert to Christianity.
See, for example, \citet{cameron}.}
provides a convenient historical dividing line.
\citet{snyder} studies pre-Constantinian pictorial art
and notes that such art is limited to four media:
frescoes, mosaics, sarcophagi, and possibly statues [Snyder (\citeyear{snyder}), page 68].
Snyder [(\citeyear{snyder}), page 87] produced a comprehensive tabulation
of pre-Constantinian Christian biblical pictorial representations.
A condensed version of Snyder's tabulation
is given here in Table~\ref{art}.\vadjust{\goodbreak}
For conciseness, we removed from Snyder's table
thirteen representations (i.e., thirteen rows)
that do not occur on sarcophagi
(and these representations happen also to be relatively rare),
and our column for ``Other media'' includes frescos and mosaics
that Snyder tabulates under separate columns.
We have, however, maintained the separate column for ``Roman fragments''
only on account of the high incidences that occur in it.
In this tabulation, note that a complete ``Jonah cycle''
had been counted as three;
also, to facilitate its reading,
we have set apart---at the top of the table---the three rows that pertain to images of Jonah.

\begin{table}
\caption{Pre-constantinian biblical pictorial
representations}\label{art}
\begin{tabular*}{\textwidth}{@{\extracolsep{\fill}}lcccc@{}}
\hline
\textbf{Biblical} & & \textbf{Roman} & \textbf{Other} & \textbf{Row} \\
\textbf{representation} & \textbf{Sarcophagi} & \textbf{fragments} & \textbf{media} & \textbf{totals} \\
\hline
Jonah cast into the sea & 8 & 23 & \phantom{0}6 & 38 \\
Jonah \& the fish & 8 & 17 & \phantom{0}3 & 28 \\
Jonah at rest & 7 & 25 & 10 & 42 \\[3pt]
Adam \& Eve & 2 & \phantom{0}0 & \phantom{0}2 & \phantom{0}4 \\
Noah in the arc & 3 & \phantom{0}2 & \phantom{0}3 & \phantom{0}8 \\
Sacrifice of Isaac & 1 & \phantom{0}2 & \phantom{0}2 & \phantom{0}5 \\
Harassment of Moses & 1 & \phantom{0}0 & \phantom{0}0 & \phantom{0}1 \\
Moses striking rock & 1 & \phantom{0}0 & \phantom{0}4 & \phantom{0}5 \\
Tobit \& fish & 1 & \phantom{0}0 & \phantom{0}0 & \phantom{0}1 \\
Daniel in lion's den & 2 & \phantom{0}0 & \phantom{0}4 & \phantom{0}6 \\
Baptism of Jesus & 1 & \phantom{0}2 & \phantom{0}3 & \phantom{0}6 \\
Jesus preaching & 1 & \phantom{0}1 & \phantom{0}0 & \phantom{0}2 \\
Healing the paralytic & 1 & \phantom{0}0 & \phantom{0}2 & \phantom{0}3 \\
Healing the possessed & 1 & \phantom{0}0 & \phantom{0}0 & \phantom{0}1 \\
Multiplying loaves \& fish & 1 & \phantom{0}1 & \phantom{0}0 & \phantom{0}2 \\
Resurrection of Lazarus & 2 & \phantom{0}1 & \phantom{0}2 &\phantom{0}5 \\
Fisherman & 2 & \phantom{0}0 & \phantom{0}1 & \phantom{0}3 \\
Woman with blood & 1 & \phantom{0}0 & \phantom{0}0 & \phantom{0}1 \\
\hline
\end{tabular*}
\end{table}

Noteworthy from this tabulation
is the very high importance placed on the story of Jonah
in pre-Constantinian pictorial representations.
Snyder [(\citeyear{snyder}), page 89] concludes from these data that:

\begin{quote}
``there can be no doubt that the primary artistic representation
of early Christianity was the Jonah cycle.''
\end{quote}

\noindent
Although the story of Jonah originates in the Old Testament,
it plays a much lesser role
in Jewish religious thought than in Christian religious thought,
with new meanings having been ascribed to it along Christian themes.
Jonah's having being spewed out by the monster fish
is symbolic of escaping death.
Snyder [(\citeyear{snyder}), page 92] points out that

\begin{quote}
``Jesus spoke of the sign of Jonah
as a prophetic paradigm of death and resurrection,
or baptism and repentance (Matthew 12:38--40).''
\end{quote}

\noindent
Jensen [(\citeyear{jensen}), page 51] states:

\begin{quote}
``Jonah, especially, serves the double function
of symbolizing both Christ's death
and his resurrection---the ``sign'' of Jonah
(Matthew 12:39 and parallels), and the baptism of each believer.''
\end{quote}

\noindent
Can it then be that the apparent ``fish'' on Ossuary \#1
is a significant, earliest known pre-cursor
to what subsequently became a
quintessentially Christian iconic representation?

The third and final items of data of which we are aware
concern the spatial distribution
of ancient tomb sites in the vicinity of Jerusalem.
We first quote from Cotton et al. [(\citeyear{cotton}), pages 8--9, by B. Isaac]:

\begin{quote}
``A dense band of rock-hewn burial caves surrounded Jerusalem on all sides,
extending to about four km from the walls of the city, the densest concentration
being closest to the walls. Most were found north, east and south of
the city.
The locations were always dictated by geology, as the graves had to be situated
where the local stone was suitable. \ldots
\ldots often the bones of more than one person were placed in the same box.
So far about 900 caves and more than 2000
ossuaries (some estimate more than 3000) have been documented.
Presumably there are many more caves that have not been discovered, and many
others were destroyed by modern construction without any record being made.
Ossuaries have been shattered or robbed by looters
or lost soon after their discovery,
and many have disappeared into private collections. Kloner and Zissu
estimate that the known caves provided burial space
for tens of thousands of people.

``The caves do not seem to form any centralized plan, but were hewn where
land was available and the rock suitable.''
\end{quote}

\begin{figure}

\includegraphics{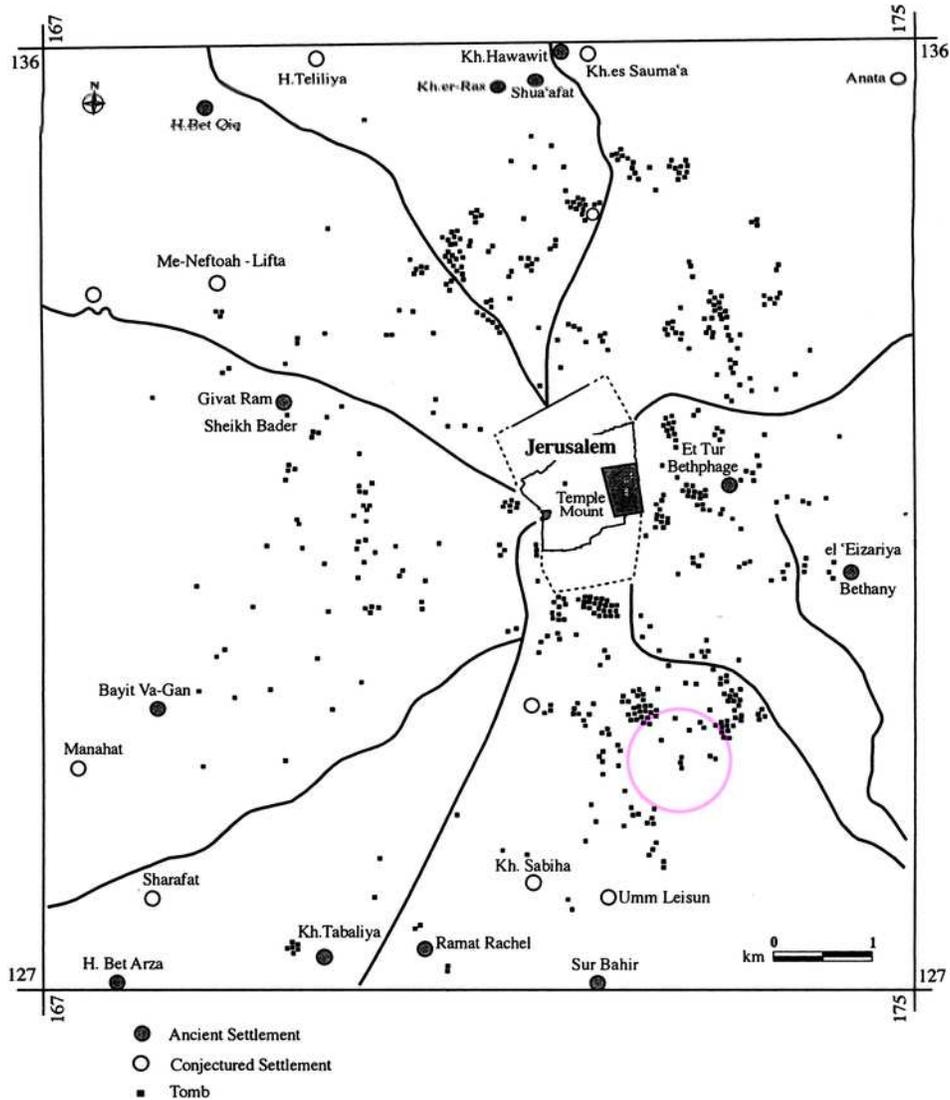}

  \caption{Map of the necropolis of Jerusalem.
[Source: Kloner and Zissu (\citeyear{klonerC}).]}\label{fig4}\vspace*{6pt}
\end{figure}

\noindent
Kloner and Zissu's (\citeyear{klonerC}) study
of the necropolis of Jerusalem contains maps
showing the locations of known tombs
and other burial areas
from the Second Temple period.
One such map, reproduced here in Figure~\ref{fig4},
covers an area of 8 $\times$ 9 kilometers---wide enough to include
some neighboring settlements.
Aside from a tendency for tombs to cluster,
the locations of tomb sites throughout Jerusalem
do not follow any particular spatial pattern.
To assist in reading this map,\
a circle of radius 500 meters has been drawn about Tomb 2.
That tomb is seen to be part of a tight cluster of three
(relatively separated from other known tomb clusters in the vicinity),
the lowest (i.e., southernmost) of which is Tomb 1,
with the one in between being the one
mentioned in Section~\ref{Sect.History}
as having been destroyed.
It is, of course, not possible to claim that this map
of tomb sites is complete,
nor that the tombs marked on it
constitute a ``simple random sample'' of all actual ones.

\section{Statistical issues: Inference} \label{Sect.Inference}

Having laid out the available sources of data,
we next consider what possibilities there are for inference
based on data of the type described.
The term ``inference,'' as used here,
has two meanings, both technical.
One is ``statistical inference,''
which typically includes producing approximations to probabilities
with more or less precisely defined inferential interpretations.
The other is ``scientific (i.e., logical) inference,'' which
is not exclusive to the domain of ``statistics''
as commonly understood.

Had a ``meaningful'' collection of names been found on the ossuaries of Tomb~2,
there might have been some possibility
of ascribing numerical weights to that new evidence
and of then combining those weights with the
numerical evidence in AF08.
For example, one among various
{a priori} \textit{candidates}
for a second tomb might (for argument's sake)
have included ``Joseph of Arimathea,''
but no such names occurred.

While a \textit{quantitative} analysis seems out of reach here,
there is nevertheless at least one
\textit{statistical principle} operating here
that is relevant under the more general rubric
of \textit{scientific inference}.
\citet{tukey} makes the well-known distinction
between \textit{exploratory} and \textit{confirmatory}
experimentation in statistics.
Within that framework, one might argue that
\textit{the data collected and analyzed
from Tomb \textup{1} played an exploratory role
in the context of designing a confirmatory experiment,
namely, that of collecting and analyzing the data from Tomb \textup{2}}.
Here, what is of essence---from a purely inferential view---is that the \textit{previously unseen} data from the second
tomb\footnote{All
that was known of Tomb 2 prior to its robotic exploration
is that it contained two ossuaries inscribed with Greek names,
and some ossuaries with incised designs;
\textit{nothing} about the designs or the inscriptions was known.}
were not obtained via exploration;
that tomb was not chosen as ``best'' of some collection
of after-the-fact examined tombs.
No:
the second tomb was selected for examination
only \textit{after} consideration had been given
to the outcome of the first experiment.
And based solely on those \textit{exploratory} considerations,
it is that \textit{one}, and only that \textit{one} tomb,
that was selected for subsequent
\textit{``confirmatory''} investigation.
Furthermore (although these are not entirely independent considerations),
not only was the second tomb chosen in this {a priori} way,
but it also has the distinction of being
the adjoining tomb---of being, literally, the tomb next door.
As such, one might argue that it provides reinforcing context
to other tombs in its immediate vicinity.
Indeed, having been involved in multiple burials
over a bounded historical period (and sharing also the rare
inscriptions ``Mara''),
it seems plausible that the families associated with these two tombs
may have had some degree of interaction and acquaintance.
In sum, the data from Tomb 2 were acquired
from an experiment of a \textit{confirmatory} nature.
The findings from the second tomb ought therefore to carry
corresponding evidentiary weight.

The main difficulty with the argument just outlined is
that the confirmatory experiment \textit{done}
was not the one that ideally \textit{needed to be done}.
The needed experiment would have been a direct test of
the null hypothesis that Tomb 1 is that
of an NT-related family, but, of course, no such test is possible.
Hence, it can be argued that the exploratory-confirmatory pardigm,
at best, applies here only partially.

What the data from the second tomb unmistakably tell us---\textit{provided one accepts some of the
interpretations posited in Section~\ref{Sect.Interpretations}}---is that this tomb was associated with a family (or families)
of relatively well-to-do individuals of Jewish origin,
some of whom had (and to no small extent)
departed from universally accepted norms and strictures of that faith,
and who apparently believed strongly enough in resurrection
to motivate a significant and deliberate
final effort on their part to express that new viewpoint.
Furthermore, this tomb---which \textit{very} likely bears connections
to Tomb 1---would therefore be one of the most (if not the most)
strikingly Judeo-Christian tomb sites ever unearthed---demonstrably more so than even 1 out of 100 tombs.
Considering how truly unique are the findings from this tomb,
and how profound are its possible meanings,
it could then be argued that some of the individuals buried there
were foremost among the earliest followers of Jesus.

Although we do not undertake
any \textit{cardinal} quantification
based on the newly acquired data,
a degree of \textit{ordinal} quantification
does, in the author's opinion, seem to be possible.
Specifically, regardless of how one chooses to quantify
the evidence from the first tomb
in respect of the likelihood
that it is or is not associated
with the New Testament family,
and regardless of how weak or how strong
one views that evidence to be,
the {a priori} evidence arising from the data of the
\textit{adjoining} tomb---provided (again) that one accepts some of the
interpretations posited in Section~\ref{Sect.Interpretations}---serves only to increase that likelihood,
and not necessarily by an entirely negligible amount.
Of course, each reader will need to decide for themselves
the plausibility of such arguments
in accordance with their assessment
of the mentioned interpretations,
with the extent to which they find
it reasonable to assume any connections between the two tombs,\footnote{One
referee has pointed out that the
characteristics of Tombs 1 and 2 are quite different.}
and with whether or not they find the partially confirmatory nature of
the experimentation
convincingly applicable to any overall analysis
of the problem being considered.

\section{Concluding remarks} \label{Sect.Conclusion}
The Rejoinder to AF08 alluded to pressures
exerted on scholars and others involved
in the work discussed here.
Having by happenstance become involved with these data,
I felt I had no choice but to pursue the facts to
their logical conclusion.
However, I did not expect new and relevant data
ever to become available.
Thus, it bears repeating that the subject
has both historical as well as archeological significance
and that the statistical issues it gives rise to have
methodological interest.
We also point out that our analyses
do not apply directly to such questions
as who was buried in any particular ossuary
or of the relationships among the individuals;
such questions necessarily entail separate inferences.
Finally, the role of coincidence,
as studied in \citet{Diaconis},
needs also to be taken into account.

Statistics is the science and the art
of quantifying and thereby reducing uncertainty,
not of eliminating it.
From a purely technical viewpoint,
the problem studied here highlights subtle aspects
of the connections between statistics
and the acts of deciding on measures of uncertainty.
Certainty itself is rarely an option.
The author is of the opinion that,
\textit{based on the currently available data},
it is at least a possibility---and one that should be considered seriously---that Tomb 1 is that of a family related to the New Testament.
This statement---not more, but also not less---stands as the author's own conclusion to the work presented here.
We must leave it to others, who may be interested,
to add to any discussions
about the relevance of statistical ideas
in assessing data of this nature.

\begin{appendix}
\section*{Appendix: Which Mary or what's in a name?} \label{Sect.Appendix}

Reliable statistical inference requires
that highly influential observations be measured reliably.
The reader will not have failed to notice
that the outcome of any analysis
to the problem considered here
is influenced heavily by a single item of data,
namely, the correct reading of the ossuary
in Tomb 1 bearing the Greek inscription.
Our aim here is not to resolve this matter for the reader,
but only to provide some context to it.

The inscription in question was shown in AF08.
It was first read,
prior to any controversies being associated with it,
by Levi Rahmani,
a foremost authority on ossuary inscriptions
whose ``eye'' for such readings has rarely been contested.
In Rahmani [(\citeyear{Rahmani}), pages~14 and~222] that inscription
is read as

\begin{quote}
``$M\alpha\rho\iota\alpha\mu\eta\nu o \upsilon (\eta)  M\alpha\rho\alpha$
of Mariamene, (who is also called) Mara\ldots

Thanks are due to the late J. Gath
for permission to publish these ossuaries\ldots

$M\alpha\rho\iota\alpha\mu\eta\nu o \upsilon$:
Here the name is the genitive of
$M\alpha\rho\iota\alpha\mu\eta\nu o \nu$,
a diminutive of
$M\alpha\rho\iota\alpha\mu\eta\nu\eta$ \ldots
one of the many variants of the name [Miriyam] \ldots
The present variant was further contracted to
$M\alpha\rho\iota\alpha\mu\nu\eta$,
which was explicitly equated with
$M\alpha\rho\iota\alpha\mu\eta$\ldots''
\end{quote}

\noindent
Rahmani goes on to say:

\begin{quote}
``($\eta$) $M\alpha\rho\alpha$:
The stroke between the \textit{upsilon} of the first and the \textit
{mu}
of the second name probably represents an \textit{eta},
standing here for the usual $\eta\ \kappa\alpha\iota$ \ldots  used
in the case of double names\ldots''
\end{quote}

\noindent
This reading of the inscription was, at the time,
corroborated by Leah di Segni
and also accepted by \citet{Kloner.three},
one of the original excavators of Tomb 1
as well as of Tomb 2.
The same reading for this inscription
is given in \citet{klonerC},
this being an English translation
of a slightly expanded version
of their earlier publication in Hebrew.
Independently of this,
in 2002 Francois Bovon, a highly respected biblical scholar,
published an article on the role of Mary Magdalene
in the \textit{Acts of Philip}
from which it might have been inferred
that Mariamne was a more likely name for Mary Magdalene
than more common variants such as Mariam [Bovon (\citeyear{Bovon})].
The plausibility of that inference is enhanced
by the fact that there are \textit{only two} other known instances
of the name version Mariamene
in all Greek of literature up to the 15th century,
both of which refer to Mary Magdalene.

Subsequent to events surrounding the publication of AF08---during which time the implications of the interaction between
Rahmani's reading
and Bovon's article became clear---two developments occurred.
First, Professor Bovon issued a clarification
through the Society of Biblical Literature
stating that he did ``not believe that Mariamne
is the real name of Mary of Magdalene'' and that
``Mariamne is, besides Maria or Mariam, a possible Greek equivalent\ldots''
Second, Rahmani's reading of the inscription
was challenged by Roger Bagnall, Stephen Pfann, Jonathan Price,
and others.
Upon reexamination, Rahmani revised his reading of the inscription.
We quote from Cotton et al.
[(\citeyear{cotton}), item 477, written by J.~Price]
which they describe as
``Ossuary of Mariam(e) with Greek inscription, 1 c. BCE - 1 c. CE''
and which they read as
``$M\alpha\rho\iota\alpha\mu\eta\ \kappa\alpha\iota\ M\alpha
\rho\alpha$,
or
$M\alpha\rho\iota\alpha\mu\ \eta \ \kappa\alpha\iota\ M\alpha
\rho\alpha$'':

\begin{quote}
``Rahmani's reading of the first name
as $M\alpha\rho\iota\alpha\mu\eta\nu o \upsilon$,
as the genitive of
$M\alpha\rho\iota\alpha\mu\eta\nu\eta/M\alpha\rho\iota\alpha\mu\nu\eta$,
has generated widespread speculation and misunderstanding.
In fact the inscribed letters are without doubt as represented here;
the mark between the \textit{iota} and last \textit{mu}
is not part of the inscription
(compare other gouges and scratches between and around the letters,
and all over the box);
the \textit{kappa} is clear (it is not an inept \textit{mu}),
and the ligature \textit{alpha-iota} is standard and unproblematic.
In a personal communication,
Rahmani has accepted the correction to his reading in the ed. pr.''
\end{quote}

\noindent
The entry goes on to say:

\begin{quote}
``The inscribed letters may be parsed in one of two ways,
without any firm criterion for preferring one or the other
(the bones in the box were not analyzed and are now reburied):
either \ldots ``Mariame and Mara''---a reading favored by SEG and BE---or \ldots ``Mariam who is also (known as) Mara''.''\footnote{The
abbreviations SEG and BE
refer to Supplementum Epigraphicum Graecum and to
Bulletin \'epigraphique in Revue des \'etudes grecques.}
\end{quote}

\noindent
While not unaware of the pressures
that must have been brought to bear,
there is nevertheless no doubt in my mind
as to the intended objectivity of these updates.

There remain three germane matters we have not yet introduced,
and which are covered
by the following three quotes from Cotton et al. (\citeyear{cotton}):

\begin{quote}
\begin{longlist}[(iii)]
\item[(i)] ``\ldots Mara is not a title, esp. not Aramaic
for ``lady'' or ``honorable woman'',
for which the correct feminine form is Marta\ldots''\footnote{It
is not within our purview to partake of such debates,
but note only that Cotton et al. do not
back up this categorical, but arguable, assertion.
[Jacobovici, private communication.]}$^,$\footnote{In
any case, if Cotton et al. are correct,
two persons are named on that ossuary.
In the Gospels, two sisters are mentioned by name:
Mary and Martha, the sisters of Lazarus.
This Mary is identified there as the one who anoints Jesus' feet
and wipes them with her hair, and has traditionally
also has been identified with Mary Magdalene.
[Jacobovici, private communication.]
This logic leads to alternate {a priori} assumptions
as in AF08.}

\item[(ii)] ``Pfann's argument that the letters KAIMARA
were added by a different hand cannot be conclusively proven,
despite the slight differences in the formation of those letters,
since in ossuary inscriptions letters are often formed
by the same inscriber in an inconsistent manner\ldots''

\item[(iii)] ``This is the only Greek inscription
recovered from the cave,
but this fact in itself is not pertinent
to the identity of the deceased,
reflecting rather the skill and choice of the inscriber.''
\end{longlist}
\end{quote}

\noindent
To these we add two observations:
If the inscription involved two hands,
then the names most likely corresponded to two different individuals.
(Mara could then, conceivably, have been a male.\footnote{It
seems, however, unlikely that any such second individual
would have been male,
not only because it is less likely for a wife
to have predeceased her husband,
but also because a husband's name
would hardly have been positioned on the ossuary
as this one's was: See Figure~1 of AF08.})
If only one hand (and at the same time) was involved,
then the inscription likely meant to identify
a person who was known by two different names
or by a title together with a name.

\begin{table}
\caption{Ossuary inscriptions with multiple names}\label{doubles}
\begin{tabular*}{\textwidth}{@{\extracolsep{\fill}}lcccc@{}}
\hline
\textbf{Genders of the names} & \textbf{Hebrew and/or} & \textbf{Greek} & \textbf{Both} & \textbf{Row} \\
\textbf{on the ossuary} & \textbf{Aramaic script} & \textbf{script} & \textbf{scripts} & \textbf{totals} \\
\hline
Father and son & \phantom{0}1 & \phantom{0}1 & 0 & \phantom{0}2 \\
Brothers & \phantom{0}3 & \phantom{0}1 & 1 & \phantom{0}5 \\
Two men; unknown & \phantom{0}3 & \phantom{0}4 & 1 &\phantom{0}8 \\[3pt]
Mother and daughter & \phantom{0}0 & \phantom{0}0 & 0 & \phantom{0}0 \\
Mother and son & \phantom{0}3 & \phantom{0}1 & 0 & \phantom{0}4 \\
Mother and children & \phantom{0}0 & \phantom{0}1 & 0 & \phantom{0}1 \\
Sisters & \phantom{0}0 & \phantom{0}1 & 0 & \phantom{0}1 \\
Two women; unknown & \phantom{0}3 & \phantom{0}2 & 1 & \phantom{0}6 \\[3pt]
Husband and wife & \phantom{0}8 & \phantom{0}2 & 6 & \phantom{0}16 \\
Brother and sister & \phantom{0}0 & \phantom{0}2 & 0 & \phantom{0}2 \\
Mixed genders; unknown & \phantom{0}0 & \phantom{0}1 & 0 &\phantom{0}1 \\[3pt]
Genders uncertain & \phantom{0}2 & \phantom{0}3 & 0 & \phantom{0}5 \\[3pt]
Column totals & 23 & 19 & 9 & 51 \\
\hline
\end{tabular*}
\end{table}

Of resulting interest are ossuaries bearing double names.
The data in Cotton et al. show that fewer than 1 in 10
among known inscribed ossuaries bore two names.
A summary of those doubly-inscribed ossuaries,
by languages used and by genders,
is provided in Table~\ref{doubles}.
The accuracy of this table is only approximate
since a few cases were either ambiguous, illegible, or both.
Also, in this table we do not distinguish
between Hebrew and Aramaic script.
Some ossuaries were, in fact, inscribed in both languages
(i.e., in Hebrew/Aramaic and in Greek).
The term ``unknown'' in three rows of the table indicates that the
relationship between the named persons could not be determined.
The Mariamne ossuary of Tomb 1
was included in this tabulation in the
``Greek script'' column, and ``Two women; unknown'' row.
Of the 51 ossuaries in this tabulation,
one other [Cotton et al. (\citeyear{cotton}), item 168, page 204, in Hebrew]
involved a Martha and Maria/Mariam.
No other among the remaining 49 double-named ossuaries
is equally noteworthy.

The inference method in AF08 is conditional
on the observed configuration of the tomb.
Here we offer only a limited observation.
If (as previously mentioned) it is assumed that Mariamne and Mara
referred to two different individuals,
and if it is assumed that both were women,
then New Testament history suggests for them a plausible
{a priori} candidate name pair,
namely, one from the general name category of Miriam/Mary
and one from the general name category of Martha/Mara;
no other two-woman name combination vies
equally for {a priori} candidacy.\footnote{A
referee suggested that the Mary and Martha of the NT
would more likely be buried in a family tomb at Bethany
with their brother,
however, women were invariably buried with their husbands.}
If we use Table~2 of AF08
and allow for the fact that
the order of the names does not matter,\footnote{The
factor of $2$ for order not mattering may
be omitted, as it occurs for all RR name assignments.}
the RR value that would then be assigned
to the actually observed pairing is\footnote{We
have not allowed here for the rarity of Mariamne
within the Miriam/Mary class, which
might arguably justify a calculation
alternative to the one shown here.}
\[
2 \times \frac{74}{317} \times \frac{21 + 7}{317}  =  \frac{1}{24.25}.
\]
We do not undertake here to consider
any required ``configurational adjustments,''
nor to carry out further calculations
based on various provisos,
but remark only that the computational consequences
that ensue are not immediately intuitive.
Of course, whatever those consequences are,
one needs also to factor in the observations from Tomb 2
and, in particular, to deal with the second Mara inscription
that appears in it.

It does seem indeed remarkable
that a question of such considerable historical interest
can sometimes revolve around the correct interpretation
of a single stroke mark on a piece of stone.
\end{appendix}

\section*{Acknowledgments}

For many helpful conversations,
for a great deal of invaluable information provided,
and for considerable other forms of assistance,
I~wish to thank
David Andrews,
James Charlesworth,
Laurel Duquette,
Felix Golubev,
Simcha Jacobovici,
Noam Kuzar,
Natalie Messika,
James Tabor,
Barbara Thomson,
and the staff at Associated Producers Ltd.
Particular thanks to SJ for bringing this data to my attention
and for facilitating many scholarly and other contacts,
and to JC and JT for generously sharing
with me their wealth of historical knowledge.
The presentation of the data made here,
and of their possible interpretations,
would not have been possible
without the benefit of extensive consultations
with SJ, JT, and JC.
However, I alone accept responsibility
for any opinions, selection of contents,
and errors in this work.
Thanks to Uitgeverij Peeters Publishers
for permission to reproduce the map in Figure~\ref{fig4}
and to Associated Producers Ltd and JT for permission
to reproduce the remaining figures appearing here.
Detailed remarks from editors and referees
have helped greatly to improve the presentation.
Finally, the author would like to retract the remark
concerning ``false memory syndrome'' made
on page 111 of the rejoinder to AF08.

%
%
%
%
%
%
%
%
%



%

\printaddresses

\end{document}